\pdfoutput=1

\newcommand\colswitch[2]{#2}

\documentclass[prd,amsmath,amssymb,floatfix,nofootinbib,superscriptaddress, \colswitch{}{twocolumn}]{revtex4}

\usepackage{bm}
\usepackage{amsmath}
\usepackage{epsf}
\usepackage{color}
\usepackage{natbib}
\usepackage{graphicx}

\def\be{\begin{equation}}
\def\ee{\end{equation}}
\def\ba{\begin{eqnarray}}
\def\ea{\end{eqnarray}}
\def\nn{\nonumber}

\def\bigoh{{\mathcal O}}
\def\ellmax{\ell_{\rm max}}
\def\n{\widehat{\bf n}}

\def\tT{\widetilde T}
\def\fnl{f_{NL}}
\def\fnlloc{f_{NL}^{\rm loc}}
\def\fnleq{f_{NL}^{\rm eq}}

\def\hfnl{\widehat f_{NL}}
\def\hB{\widehat B}
\def\tB{\widetilde B}

\def\Var{\mbox{Var}}
\def\Ee{{\mathcal E}}
\def\Nn{{\mathcal N}}

\newcommand{\threej}[6]{\left(
                           \begin{array}{ccc}
        \! #1\! & #2\!  & #3\!  \\
        \! #4\! & #5\!  & #6\!
                           \end{array}
                   \right)}

\newcommand{\sixj}[6]{\left\{
                           \begin{array}{ccc}
        \! #1\! & #2\!  & #3\!  \\
        \! #4\! & #5\!  & #6\!
                           \end{array}
                   \right\}}

 \newcommand{\vect}[1]{\bm{#1}}

\begin{document}

\title{CMB lensing and primordial non-Gaussianity}

\author{Duncan Hanson}
\affiliation{Institute of Astronomy and Kavli Institute for Cosmology Cambridge, University of Cambridge, Madingley Road, Cambridge CB3 OHA, UK}

\author{Kendrick M. Smith}
\affiliation{Institute of Astronomy  and Kavli Institute for Cosmology Cambridge, University of Cambridge, Madingley Road, Cambridge CB3 OHA, UK}
\affiliation{Department of Astrophysical Sciences, Princeton University, Princeton, New Jersey 08544, USA}

\author{Anthony Challinor}
\affiliation{Institute of Astronomy and Kavli Institute for Cosmology Cambridge, University of Cambridge, Madingley Road, Cambridge CB3 OHA, UK}
\affiliation{DAMTP, Centre for Mathematical Sciences, University of Cambridge, Wilberforce Road, Cambridge CB3 OWA, UK}

\author{Michele Liguori}
\affiliation{DAMTP, Centre for Mathematical Sciences, University of Cambridge, Wilberforce Road, Cambridge CB3 OWA, UK}

\begin{abstract}
\baselineskip 11pt
We study the effects of gravitational lensing on the estimation of non-Gaussianity from the bispectrum of the cosmic microwave background (CMB) temperature
anisotropies.
We find that the effect of lensing on the bispectrum may qualitatively be described as a smoothing of the acoustic features analogous to the temperature power spectrum. In contrast to previous results, for a Planck-like experiment which is cosmic-variance limited to $\ellmax=2000$, we find that lensing causes no significant degradation of our ability to constrain the non-Gaussianity amplitude $\fnl$ for both local and equilateral configurations, provided that the biases due to the cross correlation between the lensing potential and the integrated-Sachs-Wolfe (ISW) contribution to the CMB temperature are adequately understood. With numerical simulations, we also verify that low-order Taylor approximations to the lensed bispectrum and ISW-lensing biases are accurate.
\end{abstract}
\maketitle

\section{Introduction}

It has become clear that primordial non-Gaussianity is a powerful tool to constrain different models of inflation and shed light
on the physics of the early Universe. A large range of early-Universe models are compatible with current measurements of the CMB power spectrum, provided that they can produce small (nearly) scale-invariant primordial curvature perturbations in an otherwise flat universe. Distinguishing amongst these models will require not only additional measurements, but also characterizations of the data beyond the power spectrum. The first such statistic which is available is the bispectrum or three-point correlation function in Fourier space. As current observations already constrain the non-Gaussianity of the CMB to be weak, it can be shown that the bispectrum is also an optimal statistic to study \cite{Babich:2005en}, and so it has justifiably become the subject of much work.

The primordial bispectrum $B(\vect{k}_1,\vect{k}_2,\vect{k}_3)$ is usually characterized by an 
overall amplitude, given by the dimensionless parameter $\fnl$, and a shape specifying which 
configurations of wavevectors contain the highest contributions to the non-Gaussian signal.
Translational invariance imposes the constraint $\vect{k}_1+\vect{k}_2+\vect{k}_3 = \vect{0}$, 
and rotational and parity invariance forces the bispectrum to be a function of the lengths of the three 
wavevectors only. Thus, bispectrum shapes are often idealized as those of triangles.
The two most common choices are the local shape (hereafter often denoted as ${\rm loc}$), 
where the signal is maximum on squeezed 
configurations ($k_1 \ll k_2,k_3$); and the equilateral shape (hereafter often denoted as ${\rm eq}$),
in which the bispectrum peaks mostly on equilateral triangles ($k_1 \sim k_2 \sim k_3$). 
Many scenarios for the generation of the primordial curvature perturbation fall more or less into one of these classes.
The local shape is generally produced by models in which the perturbations are generated outside the horizon, 
curvaton models \cite{Lyth:2001nq,Enqvist:2001zp,Moroi:2008nn}.
In single-field inflation, the local shape cannot be generated at a detectable (i.e. $\gtrsim \bigoh(1)$) level;
there is a theorem which states that the single-field bispectrum in squeezed triangles is proportional to the tilt $(1-n_s)$
of the power spectrum \cite{Creminelli:2004yq}, and current observations constrain the power spectrum to be nearly 
scale-invariant \cite{Komatsu:2008hk}.
Equilateral shapes are a signature of nonstandard kinetic terms in the inflaton Lagrangian, as for example in 
DBI \cite{Alishahiha:2004eh} and ghost inflation \cite{ArkaniHamed:2003uz}.
For a complete discussion on shape classification of primordial bispectra and their correlations see \cite{Fergusson:2008ra}. 
This standard classification scheme provides a very useful interface between observation and theory.
It allows analysts to focus on constraining the amplitudes of only the fundamental bispectra, 
and it allows theorists to check rapidly whether their models are consistent with current observational constraints.

The current best 2$\sigma$ observational limits on $\fnl$ parameters from the WMAP 5-year data are \cite{Komatsu:2008hk,Smith:2009jr,Senatore:2009xx}: $-4 < \fnlloc <80$ and $-125 < \fnleq < 435$.  
Combining WMAP and SDSS data \cite{Slosar:2008hx} yields $-1 < \fnlloc < 63$. Thus the current data do not support a detection of non-Gaussianity, although the evidence for $\fnlloc$ is close to $2\sigma$. These results will soon improve dramatically: forecasted uncertainties from the Planck satellite are roughly $\sigma(\fnlloc)\approx 5$~\cite{Komatsu:2001rj}
and $\sigma(\fnleq)\approx 60$ \cite{Smith:2006ud}. Such an improvement on the present error bars should allow us to tighten significantly our present constraints on inflationary scenarios. A detection of primordial $\fnlloc \gtrsim 1$, for example, would 
rule out
standard single-field slow-roll inflation~\cite{Maldacena:2002vr,Acquaviva:2002ud}
--- a potentially sea-changing result.

Given the deep implications that a detection of primordial non-Gaussianity would have, it is crucial that all possible sources of contamination for the non-Gaussian measurement are well under control. In other words, we have to make sure that if a signal is extracted from CMB data using estimators of non-Gaussianity, it is of {\em primordial origin} and not produced by some spurious secondary or instrumental effect.

Many different sources could in principle bias a primordial non-Gaussianity measurement. 
In the analyses of WMAP data performed so far, particular attention has been devoted to 
astrophysical contaminants such as residual foreground contamination and unresolved point sources 
\cite{Komatsu:2008hk,Yadav:2007yy,Smith:2009jr} as potential spurious signals. 
Another possible source of contamination is the non-Gaussianity induced by second-order anisotropies. Beyond linear order in perturbation theory, it is no longer true that Gaussian initial conditions imply Gaussianity of the CMB temperature field. It is therefore important to study secondary anisotropies that produce non-Gaussianities of similar amplitude and shape as the primordial ones in the CMB.
In order to study this aspect in a fully consistent way, a complete numerical implementation of the second-order Einstein-Boltzmann evolution equations \cite{Bartolo:2006fj,Bartolo:2007ax,Pitrou:2007jy,Pitrou:2008hy} is necessary. Only a partial implementation is available at present \cite{Nitta:2009jp}. Meanwhile, in the absence of a full numerical solution, a number of papers on the subject \cite{Goldberg:1999xm, Smith:2006ud, Cooray:2008xz, Serra:2008wc, Khatri:2009ja, Bartolo:2008sg, Senatore:2008wk,Pitrou:2008ak} have focused on specific well-known secondaries such as gravitational lensing, the Sunyaev-Zel'dovich effect, and perturbed recombination.
%
Their effects have been found small for $\ell < 500$, and so do not form a significant source 
of contamination for WMAP. For higher resolution experiments such as Planck, however, 
they are expected to dominate over e.g. residual point sources, and must be treated with care \cite{Serra:2008wc}. 

In this paper we focus on the secondary non-Gaussianity induced by gravitational lensing of the CMB. Note that while lensing itself does not generate a three-point function, if the lensing effects are correlated to the unlensed CMB then a bispectrum may be generated. Such a correlation arises at low-$\ell$ from the integrated Sachs-Wolfe (ISW) effect or at high-$\ell$ from the nonlinear ISW
(Rees-Sciama) and Sunyaev-Zel'dovich effects. The ISW-lensing bispectrum is a direct source of bias when estimating the $\fnl$ parameters, and can ``fake'' the primordial signal if not accounted for in the analysis. This is particularly true for the local shape, as the ISW-lensing correlation sources squeezed bispectrum modes. Lensing is expected to provide the largest source of secondary bias for $\fnlloc$ estimation \cite{Serra:2008wc}. 

Apart from this direct ISW-lensing bias, the \emph{shape} of the observed bispectrum is also modified by lensing. This shape change could modify the effective normalization for a $\fnl$ estimator or even confuse the different primordial shapes with each other. In \cite{Cooray:2008xz}, for example, it was found that lensing generates a large change in the shape of the observed three-point function, degrading the experimental sensitivity to $\fnl$ (as will be discussed later, we do not reproduce this result). 

Both the ISW-lensing bias and shape change due to lensing can be approximated analytically. Gravitational lensing of the CMB is treated as a deflection of the lines of sight between the observer and recombination, with preserved surface brightness (for a recent review see \cite{Lewis:2006fu}). The lensed CMB $\tT(\n)$ is related to the unlensed CMB $T(\n)$ by
\ba
\tT(\n) &=& T[\n + \nabla\phi(\n)],   \label{eq:deflection}
\ea
where $\phi$ is the lensing potential.\footnote{%
For a discussion of the spherical displacements that are implied by
Eq.~(\ref{eq:deflection}), see Ref.~\cite{Challinor:2002cd}.
}
For analytical calculations involving $\tT(\n)$, Eq.~(\ref{eq:deflection}) is usually Taylor expanded to second order in $\phi$, and ensemble-averaged results are taken to ${\cal O}(C_{\ell}^{\phi\phi})$. The accuracy of this approximation has been studied thoroughly in the context of the lensed power spectrum \cite{Challinor:2005jy}, where it results in errors of order $10\%$ of the lensing effect at intermediate multipoles. A primary purpose of this paper is to verify with simulations of the exact lensing displacements that a low-order Taylor approximation is similarly accurate for the bispectrum.

The remainder of this paper is organized as follows. In \S\ref{sec:preliminaries} we detail our simulation and analysis steps. We calculate the
bias due to ISW-lensing in \S\ref{ssec:isw_lensing} and  we investigate the effect of lensing on the shape and normalization of the primordial bispectrum in \S\ref{ssec:lensed_b}. In \S\ref{ssec:variance} we study the increased statistical error in $\fnl$ parameters due to non-Gaussian statistics of the lensed CMB.
We summarize and draw our conclusions in \S\ref{ssec:discussion}.
The appendix provides more details of our simulation methodology.

\section{Preliminaries and Notation}
\label{sec:preliminaries}

The angular bispectrum $B_{\ell_1\ell_2\ell_3}$ is defined by
\be
\left\langle a_{\ell_1m_1} a_{\ell_2m_2} a_{\ell_3m_3} \right\rangle =
 B_{\ell_1\ell_2\ell_3}
 \threej{\ell_1}{\ell_2}{\ell_3}{m_1}{m_2}{m_3}.
\ee
Here, the $a_{\ell m}$ are the spherical-multipole coefficients
of the observed CMB and the ensemble average is taken over realizations of the primordial perturbations.
This is the most general form of the three-point function which is rotationally invariant. Under the additional assumption of parity invariance (so that $B_{\ell_1\ell_2\ell_3}=0$ if $\ell_1+\ell_2+\ell_3$ is odd and so it is invariant under all permutations) we can define the reduced bispectrum $b_{\ell_1\ell_2\ell_3}$ by
\be
B_{\ell_1\ell_2\ell_3} =
 \sqrt{ \frac{(2\ell_1+1)(2\ell_2+1)(2\ell_3+1)}{4\pi} }
 \threej{\ell_1}{\ell_2}{\ell_3}{0}{0}{0}
 b_{\ell_1\ell_2\ell_3}.
\ee
The reduced bispectra for the local and equilateral shapes can be computed efficiently as integrals involving the CMB transfer functions and the primordial power spectrum; see, for example,~\cite{Creminelli:2005hu} for details. The CMB bispectra are characterized by an amplitude- $\fnl^X$, where $X$ denotes either
local or equilateral, and a shape. We shall generally denote the
primordial CMB bispectra with unit $\fnl$ (i.e.\ the shape part) by
$B^X_{\ell_1 \ell_2 \ell_3}$.

In the limit of weak non-Gaussianity,
the minimum-variance full-sky estimator for $\fnl^X$ given cosmic-variance limited data to $\ell_{\rm max}$ is given by
\ba
\hfnl^X[a_{\ell m}] &=&  \frac{1}{6}\frac{1}{F(B^X,B^X)}
\sum_{\ell_i = 2}^{\ell_{\rm max}}
\sum_{|m_i|\leq \ell_i} \Bigg[
  B_{\ell_1\ell_2\ell_3}^X 
  \colswitch{}{\nn \\ &&}\times
\threej{\ell_1}{\ell_2}{\ell_3}{m_1}{m_2}{m_3} 
  \frac{a_{\ell_1m_1}a_{\ell_2m_2}a_{\ell_3m_3}}{C_{\ell_1}C_{\ell_2}C_{\ell_3}}
\Bigg],
  \label{eq:fnl_estimator_harmonic}
\ea
where the Fisher-matrix element $F(B,B')$ is defined for bispectra $B,B'$ by
\be
F(B,B') = \frac{1}{6} 
\sum_{\ell_1\ell_2\ell_3}^{\ell_{\rm max}}
 \frac{B_{\ell_1\ell_2\ell_3} B'_{\ell_1\ell_2\ell_3}}{C_{\ell_1}C_{\ell_2}C_{\ell_3}}.
\ee
Note that $F^{-1}(B^X,B^X)$ gives the variance of the error in $\fnl^X$ in the
Gaussian approximation.
The harmonic-space form of the estimator in Eq.~(\ref{eq:fnl_estimator_harmonic}) is too slow for practical use, but there is a mathematically equivalent, fast position-space form for the local and equilateral shapes \cite{Komatsu:2003iq,Creminelli:2005hu}.

We use non-Gaussian CMB simulations both to verify the accuracy of our low-order analytical results and to approximate quantities which are too intensive to calculate directly. Our simulations are composed of ``pairs'' of $a_{\ell m}$s:
\be
a_{\ell m} = a_{\ell m}^G + \fnl^X a_{\ell m}^{NG}.  \label{eq:ng_sim}
\ee
 The Gaussian part $a_{\ell m}^G$ sets the power spectrum of our simulations, while the non-Gaussian part $a_{\ell m}^{NG}$ sets the shape of the bispectrum. We generate the non-Gaussian part from $a_{\ell m}^{G}$ using the simulation algorithm from \cite{Smith:2006ud}. This approach is designed for maximum generality and allows us to construct non-Gaussian simulations for any bispectrum, although for specific bispectra it often contains freedoms which may be adjusted to modify the variance of the realized $a_{\ell m}^{NG}$. This is discussed further in the appendix.

Our simulations are of a flat $\Lambda$CDM cosmology with $\{ \Omega_b, \Omega_c, h, n_s, \tau, A_s \} = \{ 0.05, 0.23, 0.7, 0.96, 0.08, 2.4\times 10^{-9} \}$. We simulate the lensing potential $\phi(\n)$ as a Gaussian field which is correlated to the unlensed CMB via the ISW effect. The auto power spectra $C_{\ell}^{TT}$, $C_{\ell}^{\phi\phi}$ and cross spectrum $C_\ell^{T\phi}$ are computed using {\textsc CAMB} \cite{Lewis:1999bs}. We simulate the deflection operation of Eq.~(\ref{eq:deflection}) using the public {\textsc LENSPIX} code \cite{Lewis:2005tp, Hamimeche:2008ai}, which performs cubic interpolation on a high-resolution map. {\textsc LENSPIX} produces results which are consistent with the ``true'' lensed power spectrum calculated following \cite{Challinor:2005jy} to $0.1\%$ at $\ell < 2000$.

For our $\hfnl$ analysis, we will restrict the sum of Eq.~(\ref{eq:fnl_estimator_harmonic}) to $\ellmax=2000$ to mimic the cosmic-variance limit expected from the Planck satellite. For lensed and unlensed simulations, we use lensed and unlensed power spectra respectively in the denominator of the estimator
\cite{Babich:2004yc}.

\section{Results}

\subsection{ISW-lensing bias}
\label{ssec:isw_lensing}

The most worrying source of contamination for an analysis of primordial non-Gaussianity is the nonzero bispectrum generated by
lensing, since this can directly bias the $\fnl$ estimators. If the lens potential $\phi$ and the unlensed CMB are statistically independent, then lensing cannot generate a bispectrum, because there is a $T\rightarrow (-T)$ symmetry. This argument does not make any approximations (such as expanding to a finite order in powers of the lens potential, or assuming that the lens potential is a Gaussian field). Because there is an ISW cross correlation, however, lensing can generate a bispectrum. We are interested in the bias $\langle \hfnl^X \rangle_{\text{ISW-lensing}}$ to the $\fnl$ estimators. The ISW-lensing bispectrum can also be used as a source of cosmological information \cite{Seljak:1998nu,Spergel:1999xn,Goldberg:1999xm}, but here we concentrate on the bias to the primordial amplitudes.

\begin{figure}
\begin{center}
\includegraphics[width=\colswitch{4in}{8.5truecm}]{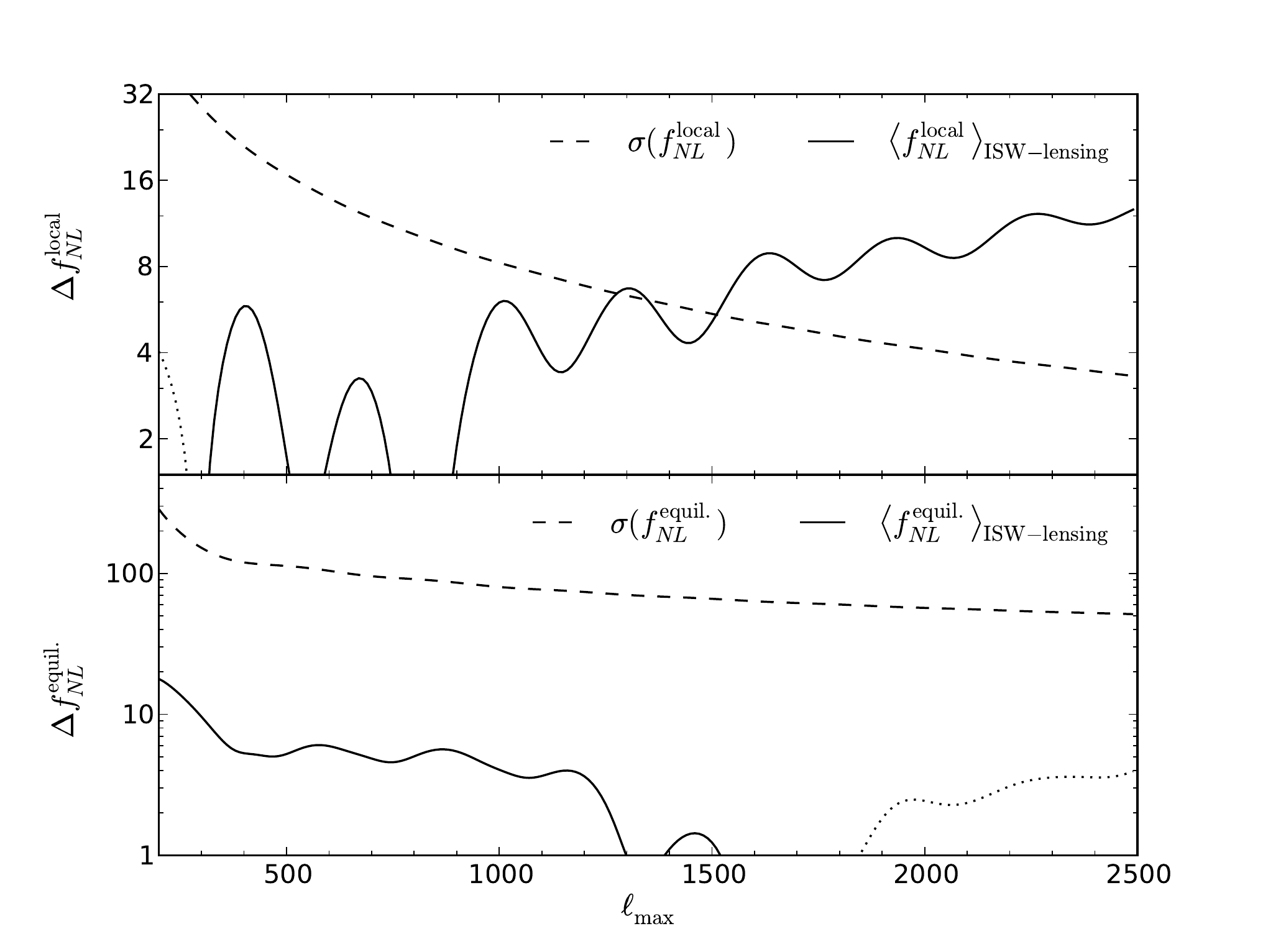}
\end{center}
\caption{Biases in $\fnl^X$
for the local (top) and equilateral (bottom) shapes if the ISW-lensing cross correlation were to be ignored. The analysis is assumed cosmic-variance limited
up to a maximum multipole $\ellmax$.
The solid/dotted lines
are calculated from Eq.~(\ref{eq:isw_lensing_lowest_order}) and are shown
dotted where the bias is negative. Long-dashed lines are the expected Gaussian
errors on $\fnl^X$ computed from the Fisher matrix.}
\label{fig:isw_fnl_biases}
\end{figure}

To lowest order in $C_\ell^{T\phi}$, we can easily predict the bias. The ISW-lensing bispectrum is
\be
B^{\text{(ISW-lensing)}}_{\ell_1\ell_2\ell_3} = f_{\ell_1\ell_2\ell_3} C_{\ell_2}^{T\phi} C_{\ell_3}^{TT} + \text{5 perm.},   \label{eq:isw_lensing_bispec}
\ee
where
\ba
f_{\ell\ell'\ell''} &=& \frac{1}{2} \left[ -\ell(\ell+1) + \ell'(\ell'+1) + \ell''(\ell''+1) \right] 
\colswitch{}{\nn \\ &&}\hspace{-0.5cm}\times
 \sqrt{\frac{(2\ell+1)(2\ell'+1)(2\ell''+1)}{4\pi}} \threej{\ell}{\ell'}{\ell''}{0}{0}{0}.
\label{eq:f}
\ea
The resultant bias to $\hfnl^X$ 
can be obtained by computing the expectation value of the estimator in Eq.~(\ref{eq:fnl_estimator_harmonic}) in the presence
of the approximate ISW-lensing bispectrum [Eq.~(\ref{eq:isw_lensing_bispec})].  A short calculation shows that
\be
\left\langle \hfnl^X \right\rangle_{\text{ISW-lensing}} = \frac{F(B^X,B^{\text{(ISW-lensing)}})}{F(B^X,B^X)}.   \label{eq:isw_lensing_lowest_order}
\ee
This bias is plotted in Fig.~\ref{fig:isw_fnl_biases} for both the local and equilateral shapes as a function of $\ellmax$ for a cosmic-variance limited $\fnl$ estimator. For local configurations, the bias is significant, but for equilateral configurations it is always at most one order of magnitude below the estimator variance, as has been observed elsewhere \cite{Smith:2006ud, Serra:2008wc}.
This behavior follows since large-scale potential fluctuations source the ISW
effect and also lens the CMB on small scales, producing a bispectrum in squeezed
triangles which is correlated with the local shape.

\begin{table}
\begin{center}
\begin{tabular}{ l c c }
\hline
\hline
& Fisher & Simulations \\
\hline
Local & $+9.3$ & $+9.4 \pm 0.2$ \\
Equilateral & $-2.4$ & $-3.1 \pm 1.8$ \\
\hline
\hline
\end{tabular}
\end{center}
\caption{Biases in $\fnl^X$ from the ISW-lensing cross correlation.}
\label{tab:1}
\end{table}

Equation~(\ref{eq:isw_lensing_bispec}) is only a leading-order approximation, when the lensing operation in Eq.~(\ref{eq:deflection}) is expanded in powers of $\phi$.
We can use the lensed simulations described in \S\ref{sec:preliminaries} to test the accuracy of this approximation
when computing biases in $\fnl$.
Table~\ref{tab:1} compares the lowest-order approximate results with those
from 100 Monte-Carlo simulations for an experiment that is cosmic-variance
limited to $\ellmax=2000$. (The errors quoted for the simulation results are
the standard error in the mean $\fnl$ from 100 simulations.)
The agreement is excellent.
As a technical point, to reduce the measurement error from the finite Monte-Carlo sample, we have subtracted the spurious contribution to $\fnl$ from the unlensed CMB
in each Monte-Carlo realization, i.e. we estimate the ISW-lensing bias as follows:
\be
\left\langle \hfnl^X \right\rangle_{\text{ISW-lensing}} =
 \Bigg\langle \hfnl^X[a_{\ell m}^{\rm lensed}] - \hfnl^X[a_{\ell m}^{\rm unlensed}] \Bigg\rangle .  \label{eq:isw_lensing_trick}
\ee
The second term on the right-hand side has zero mean since the unlensed CMB has
a vanishing three-point function,
but including it improves the statistical error on $\langle \hfnl^X \rangle_{\text{ISW-lensing}}$ due
to the finite Monte-Carlo sample.

\subsection{Lensing of the primordial bispectrum}
\label{ssec:lensed_b}

Gravitational lensing can also change the ``shape'' of the primordial bispectrum. This is a milder form of contamination than the ISW-lensing signal in \S\ref{ssec:isw_lensing}. It cannot fake a primordial signal, but it can confuse the different shapes with each other or degrade the sensitivity of the estimator.

To first order in $C_\ell^{\phi\phi}$, the lensed bispectrum $\tB_{\ell_1\ell_2\ell_3}$ is given by \cite{Cooray:2008xz}
\be
\tB_{\ell_1\ell_2\ell_3}^X = B_{\ell_1\ell_2\ell_3}^X 
 + \beta_{\ell_1\ell_2\ell_3}^X 
 + \beta_{\ell_2\ell_3\ell_1}^X 
 + \beta_{\ell_3\ell_1\ell_2}^X , \label{eq:first_order_lensing}
\ee
where $B_{\ell_1\ell_2\ell_3}$ denotes the unlensed bispectrum, and we have defined
\ba
\beta_{\ell_1\ell_2\ell_3}^X 
  &=& -\frac{\ell_1(\ell_1+1)}{4} \left( \sum_\ell \frac{\ell(\ell+1)(2\ell+1)}{4\pi} C_\ell^{\phi\phi} \right) B_{\ell_1\ell_2\ell_3}^X 
\colswitch{+}{\nn \\ &&+ \Bigg[} 
  \sum_{\ell_1'\ell_2'\ell_3'} (-1)^{\ell_1'+\ell_2'+\ell_3'} 
  \sixj{\ell_1}{\ell_2}{\ell_3}{\ell_1'}{\ell_2'}{\ell_3'}
  \colswitch{}{\nn \\ && \qquad \qquad}\times
    f_{\ell_2 \ell_1' \ell_3'} f_{\ell_3 \ell_1' \ell_2'} C_{\ell_1'}^{\phi\phi} B^X_{\ell_1 \ell_2' \ell_3'}
    \colswitch{}{\Bigg]}.
\label{eq:beta_def}
\ea
Here, the symbol in braces is a Wigner-$6j$ symbol. It is straightforward
to show that the lensed bispectrum inherits the symmetry properties of the
unlensed bispectrum, as required, since lensing respects rotational and parity
invariance.

\begin{figure}
\begin{center}
\includegraphics[width=\colswitch{4in}{8.5truecm}]{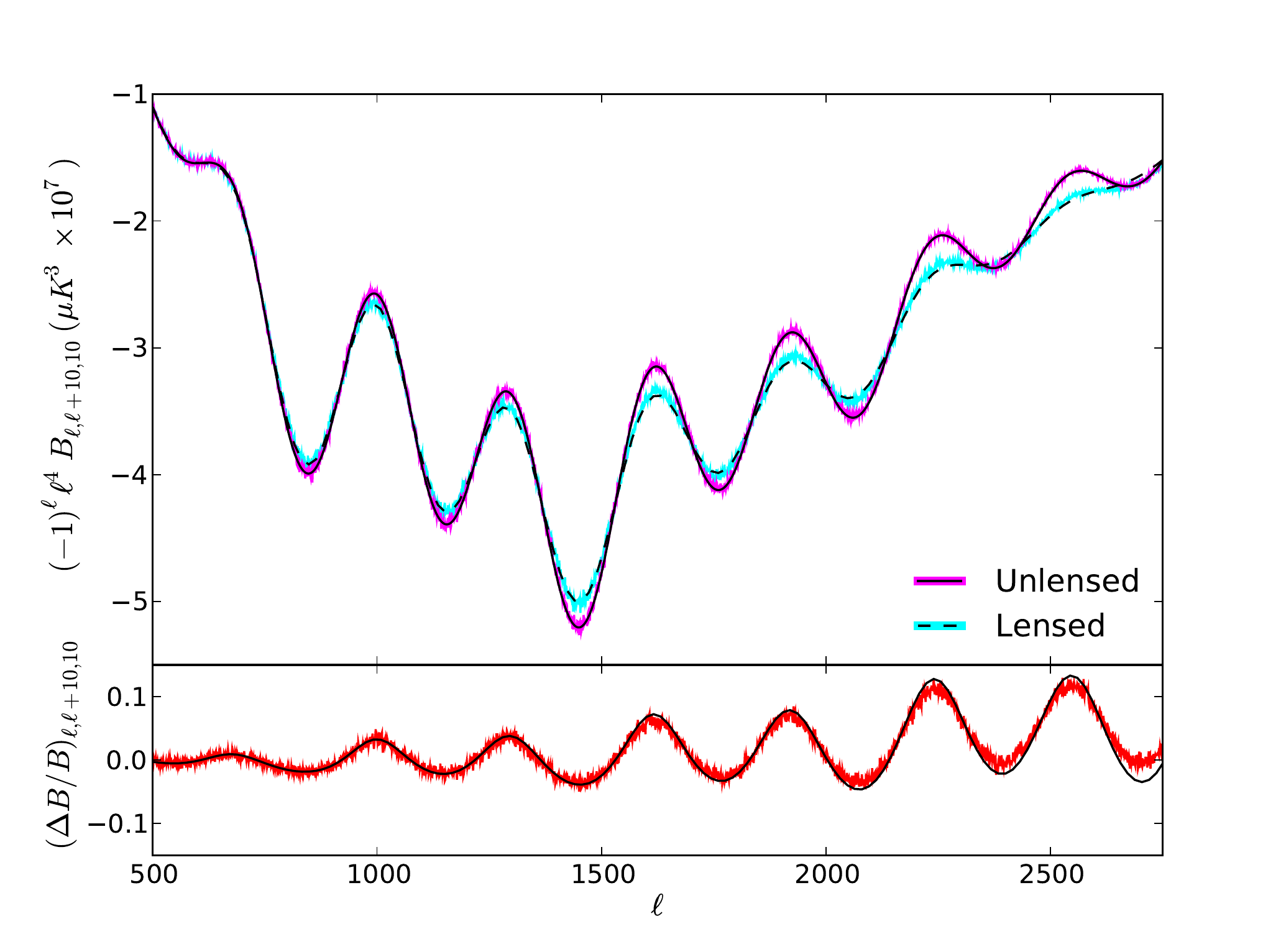}
\end{center}
\caption{A ``slice'' $B_{\ell,\ell+10,10}$ through the local bispectrum for $\fnl=1$. The simulations are unlensed (magenta line) and lensed (cyan line) with $C_{\ell}^{T\phi} = 0$, while the first-order analytical predictions are solid black (unlensed) and dashed black (lensed) lines. The Monte-Carlo results use 1000 simulations.
The fractional effect due to lensing is shown in the bottom panel for the simulations (red line) and the first-order analytic result (black line).}
\label{fig:l10_bispec_slice}
\end{figure}

Evaluating Eq.~(\ref{eq:first_order_lensing}) for both the local and equilateral shapes, we find that the difference between the lensed bispectrum $\tB_{\ell_1\ell_2\ell_3}$ and the unlensed bispectrum $B_{\ell_1\ell_2\ell_3}$ is small. This disagrees substantially with the findings of \cite{Cooray:2008xz}, although our analytical approach is the same.
In Fig.~\ref{fig:l10_bispec_slice} we plot the lensing effects on a slice through the local bispectrum. 
The agreement between our simulations and analytical results is excellent, with only small discrepancies at high-$\ell$ due to higher-order terms in $C_\ell^{\phi\phi}$ that are neglected in the analytic calculation.
The main effect of lensing on the bispectrum is a smoothing of its acoustic features, analogous to the lensing corrections to the $TT$ and $EE$ power spectra. The magnitude of the effect is also quantitatively very similar to the power spectrum case, on the order of $10\%$.

The agreement between the $\mathcal{O}(C_\ell^{\phi\phi})$ result for the
lensed bispectrum and our simulations is rather better than a casual
inspection of Eq.~(\ref{eq:beta_def}) might suggest. The first term
arises from three-point correlations of the form
\ba
\colswitch{}{&&}
\frac{1}{2} \left\langle \left( \nabla_i \phi(\n_1) \right) \left( \nabla_j \phi(\n_1) \right)
\left( \nabla^i\nabla^j T(\n_1) \right) T(\n_2)T(\n_3)  \right\rangle 
\colswitch{}{\nn \\ &&}
= \frac{1}{4} \langle \bm{\alpha}^2\rangle
\langle \nabla^2 T(\n_1)T(\n_2)  T(\n_3) \rangle
, \label{eq:threepoint1}
\ea
i.e.\ the unlensed CMB at two points
correlated with the second-order term in the Taylor expansion of the lensed
CMB at a third point. Here, 
$\langle \bm{\alpha}^2\rangle \sim (2.7\,\mathrm{arcmin})^2$
is the mean-square of the lensing
deflection angle ($\bm{\alpha} = \bm{\nabla} \phi$).
Including higher-order terms in the Taylor expansion
of the lensing action, the first term in Eq.~(\ref{eq:beta_def})
therefore generalizes
to $[\exp(-\ell_1^2 \langle \bm{\alpha}^2\rangle / 4)-1]B^X_{\ell_1 \ell_2 \ell_3}$ for $\ell \gg 1$.
This is
poorly approximated by an expansion to $\mathcal{O}(C_\ell^{\phi\phi})$
for $\ell > 1000$ and is symptomatic of poor converge of the Taylor expansion
of the lensing action \emph{in the map} on small scales. However, what matters
for the statistics of the lensed CMB is the relative displacement of points
and this is much better approximated by a low-order Taylor expansion than the
absolute displacements. When calculating the lensed
$N$-point functions in Fourier space, the importance of only relative
displacements is hidden in a near cancellation between (large) terms of the same
order in $C_\ell^{\phi\phi}$. For example, the second term in
Eq.~(\ref{eq:beta_def}) and its cyclic permutations, which arise from
correlations of the form
\be
\left\langle \left[ \bm{\alpha}(\n_1)\cdot \bm{\nabla} T(\n_1) \right]
\left[ \bm{\alpha}(\n_2)\cdot \bm{\nabla} T(\n_2) \right]
T(\n_3) \right\rangle \, ,
\ee
cancels with Eq.~(\ref{eq:threepoint1}) and its cyclic permutations in the
flat-sky limit if the three points are closely separated relative to the
correlation length of the lensing deflection. To calculate terms to
higher order in $C_\ell^{\phi\phi}$ is likely best done via the real-space
three-point function, extending the two-point methods of
Ref.~\cite{Challinor:2005jy}. However, a more accurate calculation of the
lensed bispectrum seems unnecessary given the already small effect of
$\mathcal{O}(C_\ell^{\phi\phi})$ shape changes by lensing on $\fnl$ constraints
(see below).

We now turn to the effects of lensing on the estimation of $\fnl$.
This involves an aggregate projection of the observed bispectrum onto our bispectrum of interest. In this case, the shape changes induced by lensing have the potential effectively to modify the normalization of the estimator. This may be investigated analytically if one has the capability to calculate a complete set of bispectrum coefficients, however the computational requirements of this are daunting, scaling naively as $\bigoh(\ell_{\rm max}^6)$. Rather than perform a direct calculation, we place limits on any modification to the normalization using our simulations.
We find that the estimator normalization in the presence of lensing is given by $\langle \hfnl^{\rm local} \rangle = (1.0005) \fnlloc$ for the local shape
and $\langle \hfnl^{\rm eq} \rangle = (0.965) \fnleq$ for the equilateral shape.
Both of these small modifications will only become relevant if a high-significance detection of non-Gaussianity is made for which their effect becomes
comparable to the statistical errors.
The smoothing of the bispectrum which lensing induces does not strongly affect estimators such as $\hfnl$ which average over $\ell$. We therefore conclude that
the shape changes due to lensing have a negligible effect on the normalization
of both current and near-future estimates of $\fnl$. Note, however, that the lensed CMB power spectrum should be used in the weights in the estimator
[Eq.~(\ref{eq:fnl_estimator_harmonic})] to maintain near optimality \cite{Komatsu:2001rj,Babich:2004yc}.
As CMB experiments push to higher multipoles, we expect that the additional power which lensing generates on small scales will begin to have more noticeable effects on the $\hfnl$ normalization (although other sources of non-Gaussianity will also grow rapidly). For Planck, however, the non-Gaussian effect of lensing on the estimator normalization is small enough that it may be ignored.

As a further technical point, when estimating the individual bispectrum components $B^X_{\ell_1\ell_2\ell_3}$ from the Monte-Carlo simulations in this section,
we reduced the measurement error from the Monte-Carlo samples by using the following estimator:
\ba
&&\hB_{\ell_1\ell_2\ell_3}[a_{\ell m}] =
\colswitch{ \left\langle }{\Bigg\langle}
 \sum_{m_1m_2m_3} \threej{\ell_1}{\ell_2}{\ell_3}{m_1}{m_2}{m_3} 
 \colswitch{}{\nn \\ &&} \times
 \left[ a_{\ell_1m_1} a_{\ell_2m_2} a_{\ell_3m_3} - a^G_{\ell_1m_1} a^G_{\ell_2m_2} a^G_{\ell_3m_3} \right]
\colswitch{ \right\rangle }{\Bigg\rangle} ,
\label{eq:Btrick}
\ea
where $a_{\ell m}$ denotes a Monte-Carlo simulation with $\fnl^X=1$ and $a_{\ell m}^G$ denotes the Gaussian part of the simulated temperature field (with $\fnl^X=0$). The angle brackets in Eq.~(\ref{eq:Btrick}) denote a Monte-Carlo average.
For more details on the non-Gaussian simulations, see the appendix. Subtracting the Gaussian part is analogous to the trick used for the ISW-lensing bias in \S\ref{ssec:isw_lensing} [Eq.~(\ref{eq:isw_lensing_trick})]:
the second term on the right-hand side of Eq.~(\ref{eq:Btrick})
has zero mean but improves the variance of the estimator.
Similarly, we estimate the change in normalization of the estimator $\hfnl$ due to lensing by using the estimator
\be
\Delta_{\rm lensing} \hat{f}_{NL} = \Bigg\langle
   \hfnl^{X}[a^{\rm lensed}_{\ell m}] - \hfnl^{X}[a^{\rm unlensed}_{\ell m}] \Bigg\rangle ,
\ee
where $a_{\ell m}$ denotes a Monte-Carlo simulation with $\fnl^X=1$ and no ISW-lensing correlation (i.e. $C_\ell^{T\phi}=0$).

\subsection{Lensing-induced variance}
\label{ssec:variance}

For analytical calculations of prospective $\hfnl$ sensitivity, the CMB is usually assumed to be observed on the full 
sky and perfectly Gaussian.  In this case direct calculation shows that the variance of the estimator in Eq.~(\ref{eq:fnl_estimator_harmonic})
is given by:
\be
\Var(\hfnl^X) = \frac{1}{F(B^X,B^X)} \qquad\mbox{(Gaussian statistics)}.  \label{eq:gaussian_variance}
\ee
(Note that we have assumed that each bispectrum $B^X$ is estimated independently; if multiple bispectra are being estimated jointly
then the variance would be given by $F^{-1}(B^X,B^X)$, where $F^{-1}$ is the matrix inverse.)

However, because lensing generates non-Gaussianity even if the initial conditions are Gaussian, the true variance may be larger than the right-hand side of Eq.~(\ref{eq:gaussian_variance}) due to the additional connected three-, four- and six-point
terms that are introduced. This issue has been studied for power spectrum estimation, where the excess estimator variance is small for $\ellmax=2500$ in temperature \cite{Cooray:2002fy,Smith:2006nk}, although it can be important when estimating the $B$-mode power spectrum in polarization \cite{Smith:2004up,Smith:2005ue,Li:2006pu}.

\begin{table}
\begin{center}
\begin{tabular}{ l c c c }
\hline
\hline
$\Var(\hfnl^X)$ & Fisher & Sim.\ no ISW & Sim.\ with ISW \\
\hline
Local & $17.0$ & $18.7 \pm 1.9$ & $19.7 \pm 1.9$ \\
Equilateral & $3240$ & $3710 \pm 430$ & $3720 \pm 450$ \\
\hline
\hline
\end{tabular}
\end{center}
\caption{Variance of $\fnl$ estimates in the absence of primordial
non-Gaussianity for the local and equilateral shapes.}
\label{tab:2}
\end{table}

Using our lensed simulations from the previous sections with the primordial
$\fnl=0$,
we may fit the distribution of estimated $\fnl$ values assuming that $\hfnl$ itself is approximately Gaussian distributed. We do this both for simulations with the fiducial $C_{\ell}^{T\phi}$, as well as for simulations with $C_{\ell}^{T\phi}=0$, to investigate how the cosmic variance of the ISW-lensing correlation affects the estimator variance. For the simulations with nonzero $C_{\ell}^{T\phi}$, we subtract the analytically calculated ISW-lensing bias from the $\fnl$ estimates. Our findings are given in Table~\ref{tab:2}.
For both the local and equilateral configurations, there is marginal evidence for an increased $\hfnl$ variance due to the non-Gaussianity induced by lensing. This small increase in the variance is not enough to affect significantly current Fisher estimates. In a realistic analysis of non-Gaussianity with a sky-cut and inhomogeneous noise, errors on $\hfnl$ are usually estimated from Monte-Carlo simulations with $\fnl=0$, and this increased variance will be incorporated automatically provided that the simulations are correctly lensed. 

\section{Discussion}
\label{ssec:discussion}

In the presence of lensing, the expectation value of the estimator $\hfnl^X$ can be written in the heuristic form
\be
\left\langle \hfnl^X \right\rangle = \fnl^X + \bigoh( C_\ell^{T\phi} ) + \bigoh( \fnl^X C_\ell^{\phi\phi} ).  \label{eq:conc1}
\ee
Here, $\bigoh( C_\ell^{T\phi} )$ denotes a term proportional to the cross spectrum $C_\ell^{T\phi}$, and is the ISW-lensing bias studied in \S\ref{ssec:isw_lensing}. Because the ISW-lensing bias has a negligible dependence on cosmological parameters within their currently allowed regions,
for the purposes of an analysis of primordial non-Gaussianity one can simply subtract the bias without changing the definition of the estimator $\hfnl$, provided that the bias has been reliably computed. We find that the ISW-lensing bias is quite accurately approximated by the lowest-order approximation in Eq.~(\ref{eq:isw_lensing_lowest_order}).

The $\bigoh( \fnl^X C_\ell^{\phi\phi} )$ term in Eq.~(\ref{eq:conc1}) is due to the change in the observed bispectrum shape studied in \S\ref{ssec:lensed_b}. 
In contrast to an earlier analysis~\cite{Cooray:2008xz},
we find that the difference between the unlensed and lensed bispectra is
small, on the order of $10\%$ for $\ell < 2000$,
in both the low-order approximation of Eq.~(\ref{eq:first_order_lensing}) and in our fully lensed simulations. The shape modification has an even smaller effect on the effective normalization of both the local and equilateral $\fnl$ estimators, and should not be important for Planck given its forecasted sensitivity and assuming that any detected $\fnl$ will lie within current observational bounds.
For example, were non-Gaussianity with the local shape to be detected at
$\fnlloc = 60 \pm 5$, the bias due to neglecting the lens-induced shape change
would be only $\Delta\fnlloc\approx 0.03$. Similarly, a detection of the equilateral shape with
$\fnleq = 250 \pm 60$ would have a bias $\Delta\fnleq\approx 9$.
It is fortunate that lensing of the primordial bispectrum can be ignored since
this spoils the separability of the bispectrum of the local and equilateral
models that is key to fast $\fnl$ analyses.

Finally, the variance of the $\fnl$ estimates may be analogously written
\be
\Var( \hfnl^X ) = \frac{1}{F(B^X,B^X)} + \bigoh( C_\ell^{\phi\phi} ).  \label{eq:conc2}
\ee
The first term is the variance that would be obtained if the CMB were a Gaussian field. The second term in Eq.~(\ref{eq:conc2}) represents excess variance due to non-Gaussian statistics of the lensed CMB; our simulations suggest that it is
small for a Planck-like experiment. Therefore, current Fisher-matrix forecasts
for non-Gaussianity constraints from Planck may still be considered accurate.
Furthermore, any excess variance will be automatically incorporated into any future analysis which uses lensed simulations to determine the estimator variance in the presence of a sky-cut and inhomogeneous noise.

Gravitational lensing subtly modifies the observed CMB. The bias which the ISW-lensing correlation introduces will need to be accurately subtracted in the local model. Beyond this, however, the non-Gaussianities which lensing creates represent small enough perturbations to the observed CMB that they may be safely neglected at Planck resolution.

\section{Acknowledgments}
DH was supported by a Gates Scholarship. KMS was supported by an STFC Postdoctoral Fellowship. ML was supported by STFC.

\vfill

\bibliography{fnl_lensing}

\setcounter{secnumdepth}{-1}
\appendix

\section{Appendix: Non-Gaussian simulations}
\label{app:simulations}

In this work we generate non-Gaussian simulations using the algorithm of \cite{Smith:2006ud}.
We simulate a Gaussian realization $a_{\ell m}^G$ from the power spectrum $C_\ell$,
and the non-Gaussian part $a_{\ell m}^{NG}$ of Eq.~(\ref{eq:ng_sim}) is then generated by\footnote{%
Note that we have written the simulation algorithm in a harmonic-space form which appears to have computational cost $\bigoh(\ellmax^5)$,
but for almost all bispectra of theoretical interest, there is an equivalent position-space form with cost $\bigoh(\ellmax^3)$; see
e.g.\ Eq.~(\ref{eq:alm_sim_local}) below for the case of the local shape.
}
\be
a_{\ell m}^{NG} = \frac{1}{6} \sum_{\ell_i m_i} B^X_{\ell \ell_2 \ell_3} \threej{\ell}{\ell_2}{\ell_3}{m}{m_2}{m_3} 
  \frac{a_{\ell_2 m_2}^{G\ast}}{C_{\ell_2}} \frac{a_{\ell_3 m_3}^{G\ast}}{C_{\ell_3}}.
\label{eq:sim_bispec_alg}
\ee
Note that $a_{\ell m}$ has zero mean for $\ell \neq 0$. Since we do not simulate
the monopole it is not necessary to subtract the mean explicitly.
The above algorithm is a completely general method to create weakly non-Gaussian simulations with specified power spectrum $C_\ell$ and bispectrum $B_{\ell_1\ell_2\ell_3}$,
provided that contributions of order $\bigoh(\fnl^2)$ and higher can be neglected.
More precisely, the two-, three-, and connected $N$-point functions of the simulated field satisfy
\ba
\left\langle a_{\ell_1m_1}^* a_{\ell_2m_2} \right\rangle &=& 
   \left[ C_{\ell_1} + f_{NL}^{2}C_{\ell_1}^{NG} \right] \delta_{\ell_1\ell_2} \delta_{m_1m_2}  \colswitch{\label{eq:sim_alg_2pt}}{\nn} \\
\left\langle a_{\ell_1 m_1} a_{\ell_2 m_2} a_{\ell_3 m_3} \right\rangle &=&
   \left[ \fnl B_{\ell_1 \ell_2 \ell_3} + \bigoh\left( \fnl^3 \right) \right]
   \colswitch{}{\nn \\ && \qquad}\times
   \threej{\ell_1}{\ell_2}{\ell_3}{m_1}{m_2}{m_3}  \label{eq:sim_alg_3pt}  \colswitch{}{\nn} \\
\left\langle a_{\ell_1m_1} a_{\ell_2m_2} \cdots a_{\ell_Nm_N} \right\rangle_{\text{c}} &=&
   \bigoh(\fnl^{N-2})
, \qquad (N \ge 4) . \label{eq:sim_alg_Npt}
\ea
This simulation method is very general and computationally efficient, but one caveat is that terms of $\bigoh(\fnl^2)$ and higher
are not explicitly controlled.
For many purposes this caveat is irrelevant; to consider a concrete example, suppose we have an estimator $(\Ee/\Nn)$ of $\fnl$ whose overall normalization $\Nn$
is unknown, and we are using the simulations to determine $\Nn$.  (This approach to normalizing the estimator was used to analyze WMAP data
in \cite{Smith:2009jr,Senatore:2009xx}.)  Then we can compute $\Nn$ as the following Monte-Carlo average:
\be
\Nn = \left\langle \left( \frac{d}{d\fnl} \right)_{\fnl=0} \Ee[a_{\ell m}] \right\rangle  \label{eq:sim_alg_mc} .
\ee
In this form, it is clear that the terms containing two or more powers of $\fnl$ are irrelevant, since they will not contribute to the
derivative on the right-hand side.
However, there are cases where these terms are important; a concrete example would be the $\bigoh(\fnl^2)$ contribution to the variance,
$\Var(\hfnl)$, of the $\fnl$ estimator which was studied in \cite{Creminelli:2006gc,Liguori:2007sj}.
This term is proportional to the connected four-point function, and is not simulated reliably by the simulation algorithm in 
Eq.~(\ref{eq:sim_bispec_alg}).

\begin{figure}
\begin{center}
\includegraphics[width=\colswitch{4in}{8.5truecm}]{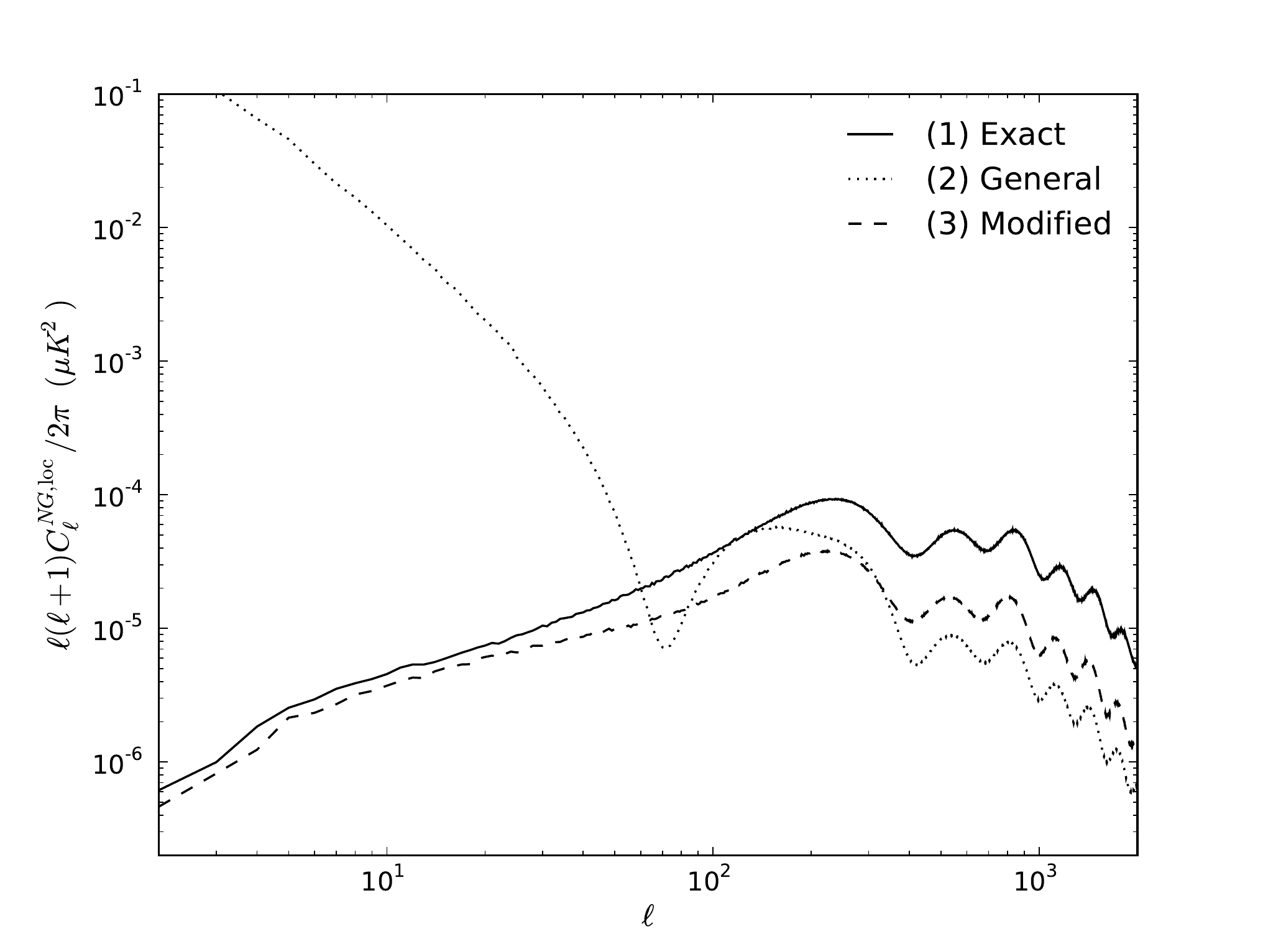}
\end{center}
\caption{Power spectrum of the non-Gaussian component $a_{\ell m}^{NG,\text{loc}}$ in the local model from three different simulation techniques:
(1) uses the ``exact'' simulation algorithm of \cite{Liguori:2003mb};
(2) uses the general algorithm of Eq.~(\ref{eq:alm_sim_local}); and (3) uses the modified algorithm of Eq.~(\ref{eq:sim_local}).}
\label{fig:sim_algorithms}
\end{figure}

For the local shape, we found that the ``subleading'' $C_{\ell}^{NG}$ contribution to the power spectrum in Eq.~(\colswitch{\ref{eq:sim_alg_2pt}}{\ref{eq:sim_alg_Npt}}) is spuriously large on large scales when using this simulation algorithm (Fig.~\ref{fig:sim_algorithms}).
It is frequently possible to work around this problem if one is aware of it.
For example, where simulations are used to determine the estimator normalization $\Nn$ [Eq.~(\ref{eq:sim_alg_mc})],
the machinery from \cite{Smith:2006ud} allows the derivative with respect to $\fnl$ to be computed exactly, with zero contribution from higher
powers of $\fnl$.
However, our preferred solution is to make a simple change to the simulation algorithm for the special case of the local shape,
which will give a reasonable power spectrum on large scales, as we now explain.

First, note that in the local model, the reduced bispectrum
can be written for $\fnlloc=1$ as
\begin{equation}
b^{\text{local}}_{\ell_1 \ell_2 \ell_3}  =   2\int dr \,r^2\,\alpha_{\ell_1}(r) \beta_{\ell_2}(r) \beta_{\ell_3}(r) + \text{cyc.}, 
\label{eq:bispec_red}
\end{equation}
where the functions $\alpha_\ell(r)$ and $\beta_\ell(r)$ involve integrals
over wavenumber $k$ of the CMB transfer functions, spherical Bessel functions,
$j_\ell(kr)$, and the primordial power spectrum; see e.g.~\cite{Creminelli:2005hu} for details. Using this in Eq.~(\ref{eq:sim_bispec_alg}) gives
\colswitch{\ba
a_{\ell m}^{NG,{\rm loc}} &=& \frac{1}{3} \int dr\, r^2 \left[
   \alpha_\ell(r) \left( \int d^2\n\, Y_{\ell m}^*(\n) B(r,\n)^2 \right)\right] 
 +
  \frac{2}{3} \int dr\, r^2 
\left[
\beta_\ell(r) 
\left( \int d^2\n\, Y_{\ell m}^*(\n) A(r,\n) B(r,\n) \right) 
\right],  \label{eq:alm_sim_local}
\ea}{ 
\ba
a_{\ell m}^{NG,{\rm loc}} &=& \int dr\, r^2 \Bigg[
  \frac{2}{3} \beta_\ell(r) \left( \int d^2\n\, Y_{\ell m}^*(\n) A(r,\n) B(r,\n) \right) 
\nonumber \\ && + 
   \frac{1}{3} \alpha_\ell(r) \left( \int d^2\n\, Y_{\ell m}^*(\n) B(r,\n)^2 \right) 
\Bigg] ,  \label{eq:alm_sim_local}
\ea } 
where we have defined
\ba
A(r,\n) &=& \sum_{\ell m} \alpha_\ell(r) a_{\ell m}^G Y_{\ell m}(\n) /C_\ell \\
B(r,\n) &=& \sum_{\ell m} \beta_\ell(r) a_{\ell m}^G Y_{\ell m}(\n) / C_\ell .
\ea
If we modify the simulation algorithm by defining
\be
a_{\ell m}^{NG,{\rm loc}'} = \int dr\, r^2 \left[\alpha_\ell(r) \left( \int d^2\n Y_{\ell m}^*(\n) B(r,\n)^2 \right)\right],  \label{eq:sim_local}
\ee
then the power spectrum on large scales is much smaller (see Fig.~\ref{fig:sim_algorithms}), while
it is easy to see that the expressions for the $N$-point functions in Eq.~(\colswitch{\ref{eq:sim_alg_2pt}}{\ref{eq:sim_alg_Npt}}) are unmodified.

For the local shape, there is a different simulation algorithm available \cite{Liguori:2003mb} which is ``exact'', in the sense 
that the $N$-point functions are reliably simulated (including subleading terms) for arbitrary $N$.
This algorithm works by simulating the initial Newtonian potential $\Phi({\bf r})$ inside a spherical shell which contains
the surface of last scattering, and is thick enough to include all points inside the causal horizon at recombination.
In Fig.~\ref{fig:sim_algorithms}, we also show the power spectrum of non-Gaussian simulations obtained using the exact
simulation algorithm.
It is seen that the general simulation algorithm [Eq.~(\ref{eq:sim_bispec_alg})] produces a large-scale power spectrum which is
much larger than the exact algorithm, but our modified algorithm for the local shape [Eq.~(\ref{eq:sim_local})] produces a
power spectrum which is smaller than the exact algorithm by an $\bigoh(1)$ factor.

To understand intuitively why the modification proposed for the local shape in Eq.~(\ref{eq:sim_local}) eliminates the
spuriously large power spectrum on large scales which is generated by the general algorithm [Eq.~(\ref{eq:sim_bispec_alg})],
we note that the modified algorithm has the following maximum-likelihood interpretation.
Starting from a Gaussian CMB realization $a_{\ell m}^G$, suppose that we ``guess'' the potential $\Phi({\bf r})$ throughout
the Hubble volume, by finding the potential which maximizes the likelihood
\be
\exp\left(-\frac{1}{2} \int \frac{d^3{\bf k}}{(2\pi)^3} \frac{|\Phi({\bf k})|^2}{P(k)}\right)
\ee
subject to the constraint that the potential $\Phi({\bf r})$ projects to the observed CMB $a_{\ell m}^G$.
Suppose we then define $a_{\ell m}^{NG}$ by projecting the potential $\Phi({\bf r})^2$ to an observed set of CMB multipoles.
A short calculation then shows that the operation $a_{\ell m}^G \rightarrow \Phi({\bf r}) \rightarrow a_{\ell m}^{NG}$
defined by this procedure is the same as the modified algorithm defined in Eq.~(\ref{eq:sim_local}).
Thus our modified algorithm for the local shape is closely related to the exact algorithm, but it produces a somewhat 
smaller power spectrum because the deprojection operation $a_{\ell m} \rightarrow \Phi({\bf r})$ only generates power in
one radial mode for each value of $\ell$.
In contrast, the general simulation algorithm in Eq.~(\ref{eq:sim_bispec_alg}) does not appear to have any direct physical
correspondence with the exact algorithm, and there is no guarantee that the power spectra are comparable.

For the equilateral shape, we find that the ${\cal O}(f_{NL}^{2})$ component of the power spectrum is reasonable on all
scales and no modified version of the general simulation algorithm seems to be necessary.  Formulating an exact simulation
algorithm for the equilateral shape (i.e. an algorithm which precisely simulates $N$-point correlations for $N\ge 4$)
has not been done; this would presumably require going back to the inflationary physics.

In summary, there are several possible non-Gaussian simulation algorithms, with trade-offs as follows.
For the equilateral shape, the general algorithm in Eq.~(\ref{eq:sim_bispec_alg}) is the only known simulation procedure,
and does not appear to contain any problems such as a spuriously large power spectrum.
For the local shape, there is an algorithm \cite{Liguori:2003mb} which is exact but computationally expensive (roughly
100 CPU hours for the resolution requirements of this paper).
This algorithm must be used for studies which require $\bigoh(\fnl^2)$ and higher contributions to the $N$-point
functions to be simulated precisely (such as the $\bigoh(\fnl^2)$ contribution to $\Var(\hfnl^{\rm loc})$ studied in
\cite{Creminelli:2006gc,Liguori:2007sj}).
Otherwise, the algorithm proposed here for the local shape [Eq.~(\ref{eq:sim_local})] provides an alternative to the exact
algorithm which is closely related but rather faster (roughly 20 CPU minutes).

\end{document}